# Substrate-induced spin-torque-like signal in spin-torque ferromagnetic resonance measurement

Dingsong Jiang[1#], Hetian Chen[2#], Guiping Ji[1], Yahong Chai[1], Chenye Zhang[1], Yuhan Liang[2], Jingchun Liu[2], Witold Skowroński[3], Pu Yu[4], Di Yi[2*], Tianxiang Nan[1*]

1. School of Integrated Circuit and Beijing National Research Center for Information Science and Technology (BNRist), Tsinghua University, Beijing China

2. School of Materials Science and Engineering, Tsinghua University, Beijing China

3. AGH University of Science and Technology, Institute of Electronics, Kraków Poland

4. Department of Physics, Tsinghua University, Beijing, China

#: The authors contribute equally to this work.

* diyi@mail.tsinghua.edu.cn, nantianxiang@mail.tsinghua.edu.cn

# Abstract

Oxide thin films and interfaces with strong spin-orbit coupling have recently shown exceptionally high charge-to-spin conversion, making them potential spin-source materials for spintronics. Epitaxial strain engineering using oxide substrates with different lattice constants and symmetries has emerged as a mean to further enhance charge-to-spin conversion. However, high relative permittivity and dielectric loss of commonly used oxide substrates, such as $SrTiO_3$, can cause significant current shunting in substrates at high frequency, which may strongly affect spin-torque measurement and potentially result in an inaccurate estimation of charge-to-spin conversion efficiency. In this study, we systematically evaluate the influence of various oxide substrates for the widely-used spin-torque ferromagnetic resonance (ST-FMR) measurement. Surprisingly, we observed substantial spin-torque signals in samples comprising only ferromagnetic metal on oxide substrates with high relative permittivity (e.g., $SrTiO_3$ and $KTaO_3$), where negligible signal should be initially expected. Notably, this unexpected signal shows a strong correlation with the capacitive reactance of oxide substrates and the leakage radio frequency (RF) current within the substrate. By revising the conventional ST-FMR analysis model, we attribute this phenomenon to a 90-degree phase difference between the RF current flowing in the metal layer and in the substrate. We suggest that extra attention should be paid during the ST-FMR measurements, as this artifact could dominate over the real spin-orbit torque signal from high-resistivity spin-source materials grown on substrate with high relative permittivity.

# I. Introduction

Magnetic random-access memories (MRAMs) and spintronic logical devices based on spin-orbit torques (SOTs) have garnered significant research attention due to their low-power consumption, high endurance, high-speed, and non-volatile properties [1-10]. The development of materials capable of generating SOT with high-efficiency and the establishment of accurate metrologies for quantifying current-induced SOTs in various material systems are crucial. To evaluate SOT efficiency of spin source materials, several experimental techniques have been devised to detect static or dynamic changes in magnetization of ferromagnet (FM) induced by SOTs [11]. These methods include spin-torque ferromagnetic resonance (ST-FMR) [12-17], second harmonic Hall voltage (SHHV) [18-20], measurements of current-induced hysteresis loop shift [21-23], measurements of current-induced magnetization switching [24], and optical measurements of current-induced effective field [25-27], etc. Among these techniques, ST-FMR stands out as a widely employed method for measuring SOTs due to its self-calibration, simplicity, and versatility across a wide range of material systems [12-14,28-31]. ST-FMR allows the examination of magnetization dynamics induced by SOTs at ferromagnetic resonance frequencies of the gigahertz range, providing insights into both damping-like and field-like SOTs. During the ST-FMR measurements, spin pumping from the magnetic layer to the spin source layer, coupled with the inverse spin-Hall effect, can contribute to the spin torque signal and is considered as the major source of artifact, which has been carefully evaluated [31,32]. On the other hand, despite considerable attention paid to the fact that substrates could become highly conductive at high frequencies and generate spin-torque-like signals, the impact of substrates to the ST-FMR measurements has not been thoroughly investigated.

Remarkably, the influence of substrates could be even more pronounced when measuring SOTs in oxide spin source materials, which have recently exhibited an exceptionally strong spin-Hall effect [28,30,33-36]. Firstly, oxide spin source materials typically have higher resistivity compared to heavy metals at room-temperature (i.e. Pt ~ 20 $\mu\Omega$cm [12], $SrIrO_3$ ~ 570 $\mu\Omega$cm [37], $SrRuO_3$ ~ 290 $\mu\Omega$cm [36]). Secondly, oxide substrates, such as $SrTiO_3$ and $KTaO_3$ which are commonly employed for epitaxial growth of oxide spin sources [15,29,30,35,36] and two-dimensional electron gas (2DEG) systems [33,38-40], exhibit high conductivity at high frequencies due to their high dielectric constant and dielectric loss [41,42]. This combination leads to a much more pronounced RF current shunting effect from the substrate in oxide spin sources as compared to heavy metal materials in ST-FMR measurements.

In this work, we investigated the artifact spin-torque signal in ST-FMR measurements on different oxide substrates with a wide range of relative permittivity. We find that the artifact spin torque signal is large in samples of only ferromagnetic metal permalloy (Py) on oxide substrates with high relative permittivity (without a spin source material). This artifact spin torque signal is inversely proportional to the capacitive reactance of substrates, which can be attributed to the Oersted field generated by an off-phased leakage RF current in oxide substrates according to our revised ST-FMR analysis model. Notably, we demonstrate that this artifact spin-torque signal can significantly impact the spin-Hall ratio analysis of Pt, a benchmarked spin-Hall source material with high conductivity, when using $SrTiO_3$ substrate. When using the more resistive oxide spin-Hall source materials such as $SrIrO_3$, the spin torque signal from $SrIrO_3$ could be overwhelmed by the artifact from $SrTiO_3$ alone. Our findings offer a deeper insight into the origin of spin torque signal measured in oxide spin-Hall sources and pave the way for a more precise measurement of SOT materials.

## II. Spin-torque ferromagnetic resonance (ST-FMR)

In ST-FMR measurements, a microwave RF current is applied on the microstrip of spin-Hall source material/FM bilayer. Due to the spin-Hall effect (SHE) [43] or Rashba-Edelstein effect (REE) [44], an RF spin current is generated in the spin-Hall source layer and inject into the adjacent FM layer to induce a magnetization precession, which leads to an oscillation of the microstrip resistance because of the anisotropic or spin-Hall magnetoresistance in the FM layer. The ST-FMR voltage signal can be detected by mixing the RF current with the oscillating resistance [12,13,45]. By solving the Landau-Lifshitz-Gilbert (LLG) equation, one can obtain the mixing voltage as a function of the external magnetic field [12,13,31], as expressed in Eq. (1):

$$V_{mix}(H_{ext}) = V_S \frac{\Delta H^2}{\Delta H^2 + (H_{ext} - H_0)^2} + V_A \frac{\Delta H (H_{ext} - H_0)}{\Delta H^2 + (H_{ext} - H_0)^2}, \quad (1)$$

where $H_{ext}$ is the external magnetic field, $\Delta H$ and $H_0$ are the linewidth and resonance field, respectively. By employing the fitting procedure of Eq. (1) consisting of symmetric and antisymmetric Lorentzian functions to the experimental data, two voltage components can be extracted: a symmetric voltage component $V_S$ proportional to the strength of the damping-like torque $\tau_{DL}$, and an antisymmetric voltage component $V_A$ proportional to the strength of the sum of field-like torque $\tau_{FL}$ and RF Oersted field $H_{RF}$, according to the Eq. (2) and Eq. (3), respectively:

$$V_S = -\frac{1}{4}\frac{dR}{d\varphi}\frac{I_{RF}\cos\varphi}{2\pi\Delta H (df/dH_{ext})|_{H_{ext}=H_0}}\tau_{DL}, \quad (2)$$

$$V_A = -\frac{1}{4}\frac{dR}{d\varphi}\frac{I_{RF}\cos\varphi\sqrt{1+(4\pi M_{eff}/H_0)}}{2\pi\Delta H (df/dH_{ext})|_{H_{ext}=H_0}}(\tau_{FL} + \gamma\mu_0 H_{RF}), \quad (3)$$

where $f$ is the excitation frequency of ST-FMR measurement, $\gamma$ is the gyromagnetic ratio, $\mu_0$ is the permeability in vacuum, $R$ is the resistance of microstrip, $I_{RF}$ is the RF current through the microstrip, $4\pi M_{eff}$ is the effective demagnetization field, $\varphi$ is the angle between $I_{RF}$ and $H_{ext}$. Finally, the spin-Hall ratio $\theta_{SH}$ can be calculated via [15,45]:

$$\theta_{SH} = \tau_{DL}\frac{2e}{\hbar}\frac{M_s t l \rho}{I_{RF} R}, \quad (4)$$

where $\hbar$ is the reduced Planck's constant, $e$ is the electron charge, $l$ is the length of the microstrip, $t$ is the thickness of FM layer, $\rho$ is the resistivity of spin-Hall source layer. In Eq. (4), the only unknown parameter is the RF current $I_{RF}$ which can be measured using a vector network analyzer [45,46]. Alternatively, in a simpler scenario, the spin-Hall ratio $\theta_{SH}$ can be estimated by considering the ratio of $V_S$ to $V_A$, denoted as

$\theta_{SH} = (V_S/V_A)\sqrt{1+4\pi M_{eff}/H_0}\frac{e\mu_0 M_s t d}{\hbar}$, assuming that $V_A$ is primarily contributed from the Oersted field, where $d$ is the thickness of spin-Hall source layer [12,29,30]. It should be noted that in systems with significant field-like torques [17,47,48], the spin-Hall ratio obtained solely from the ratio of $V_S$ to $V_A$ may lead to misestimations. Nevertheless, accurate measurements of the symmetric component ($V_S$) are of great importance for determination of the spin-Hall ratio in ST-FMR measurements.

## III. Device preparation and measurements

In this study, we used oxide substrates with different relative permittivity $\varepsilon_r$ at room temperature during the ST-FMR measurements. The substrates used were 5 mm×5 mm, 500 μm-thickness (001)-oriented SrTiO$_3$ (STO, $\varepsilon_r$ ~ 300[41]), (001)-oriented KTaO$_3$ (KTO, $\varepsilon_r$ ~ 290[42]), (001)-oriented (LaAlO)$_{0.3}$-(Sr$_2$AlTaO$_6$)$_{0.7}$ (LSAT, $\varepsilon_r$ ~ 22.6[50]), (110)-oriented DyScO$_3$ (DSO, $\varepsilon_r$ ~ 26[51]), and (0001)-oriented Al$_2$O$_3$ ($\varepsilon_r$ ~ 9.2[52]), and all substrates were not be treated with any solutions. To make the samples without any spin-Hall source materials, we deposited only the ferromagnetic thin films of Py (5 nm) capped with Ti (1.5 nm) onto these oxide substrates by using high-vacuum magnetron sputtering at an Ar pressure of 3 mTorr. To make the control samples, bilayers of Pt (5nm)/Py (5nm) and Py (5nm)/Pt (5nm) with reversed layered sequence were deposited on STO, and Pt (5nm)/Py (5nm) was deposited on Si ($\rho$ ~ 0.01 Ω·cm)/SiO$_2$ (500 nm). For samples with oxide spin-Hall-source materials, 17 nm SrIrO$_3$ (SIO) was grown on a (001)-oriented STO substrate by using a pulsed laser deposition (PLD) with a 248 nm KrF excimer laser at an oxygen pressure of 0.1 Torr. During the PLD growth, the growth temperature was 700 °C, and the laser energy density was maintained at 1.2 J/cm². Subsequently, the sample was cooled to room temperature in an oxygen atmosphere of 30 Torr. Then a 5 nm Py was deposited onto the samples by sputtering. All samples were fabricated using the lithography and ion-beam etching processes, into the 16 μm×80 μm microstripes with ground-signal-ground (GSG) electrodes for ST-FMR measurements and 16 μm×80 μm Hall bars for SHHV measurements. Especially, for the STO/Py, STO/Pt/Py and STO/Py/Pt samples, we confirmed that the STO substrates were still insulative after the device fabrication processes by measuring the dc resistance (> 200 MΩ) between two metal pads separated by ~ 200 μm.

Fig.1 shows the schematic of ST-FMR measurement setup, in which a microwave current with a power of 15 dBm is applied on the microstrip along the +x direction from a microwave generator and is modulated by a low frequency sinewave at 1713 Hz. The mixing voltage signal was measured by a lock-in amplifier as a function of the external magnetic field $H_{ext}$. To characterize the impedance of samples, we used a vector network analyzer (VNA) to measure the scattering parameter $S_{11}$ (see S1 in Supplementary Material [49] for a detailed description of the measurements).

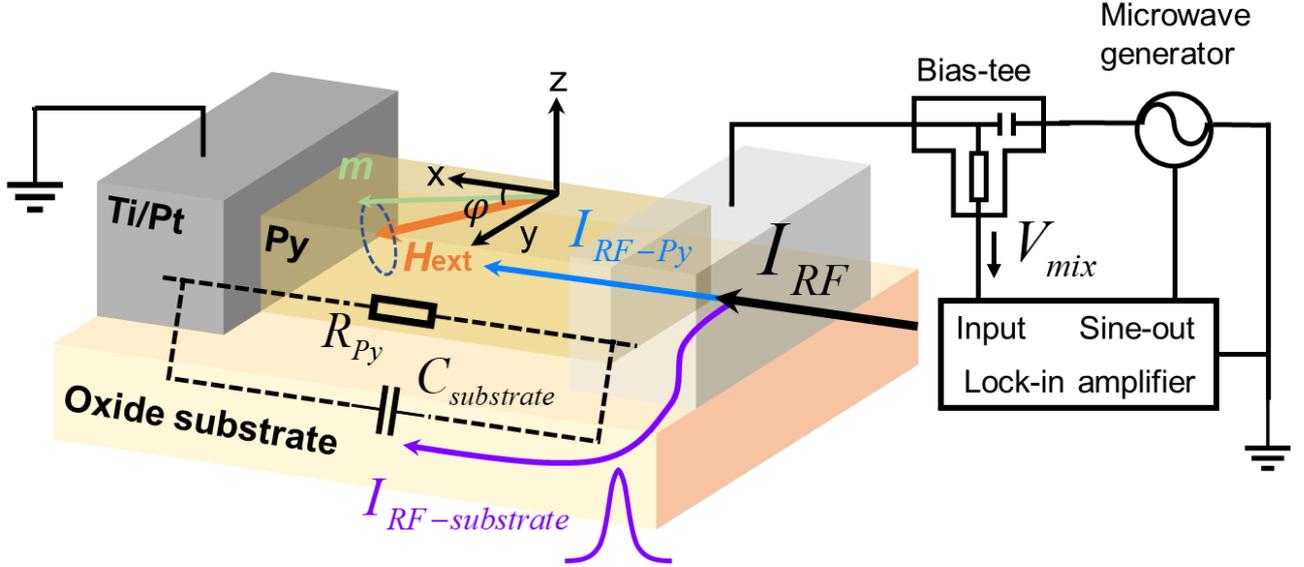

**Fig. 1.** Schematic of ST-FMR measurement set-up and RF current shunting model. $H_{ext}$ is the external magnetic field, $\varphi$ is the angle between RF current $I_{RF}$ and external magnetic field $H_{ext}$, $m$ is the magnetization of Py layer. $I_{RF}$, $I_{RF-substrate}$, $I_{RF-Py}$ are the total RF current through oxide substrate/Py device, leakage RF current in oxide substrate, and RF current in Py, respectively. $R_{Py}$ is the equivalent resistor of Py layer, and $C_{substrate}$ is the capacitor formed in oxide substrate through the ground-signal-ground (GSG) electrode.

## IV. Results

Fig.2(a)-(e) show the typical ST-FMR spectra of 5 nm Py on substrates of STO, KTO, Al$_2$O$_3$, DSO, and LSAT measured at $f$ = 5 GHz, $\varphi$ = 225°, which can be well fit by the sum of symmetric and antisymmetric Lorentzian functions expressed in Eq. (1). We found that the symmetric voltage signal that was attributed to damping-like torque appears in all five samples. Noticeably, the symmetric voltage signal from STO/Py (5nm), KTO/Py (5nm) has the same order of magnitude as Si/SiO$_2$/Pt (5nm)/Py (5nm) control sample which has a substantial damping-like torque with a spin-Hall ratio of 0.071 ± 0.002 (see S2 in Supplementary Material [49] for details). However, in all samples including STO/Py and KTO/Py, the damping-like torque should be negligible since there is no spin source layer [12], as confirmed by an independent low-frequency SHHV measurements (see S3 in Supplementary Material [49] for details of SHHV measurements). The discrepancy between ST-FMR and SHHV measurements for the oxide substrate/Py devices (especially the heterostructures of STO/Py and KTO/Py) implies that the symmetric signal from ST-FMR measurements is not originated from SOT but might be an artifact signal. Furthermore, we discovered that this artifact signal possessed the same symmetry as that of SOT, with the same angular dependence as Si/SiO$_2$/Pt/Py control sample (see Fig. S6 in Supplementary Material [49]). Additionally, small antisymmetric components are also observed from all oxide substrate/Py samples, which could be originated from a field-like torque due to the Rashba effect at the Py/Ti interface [53].

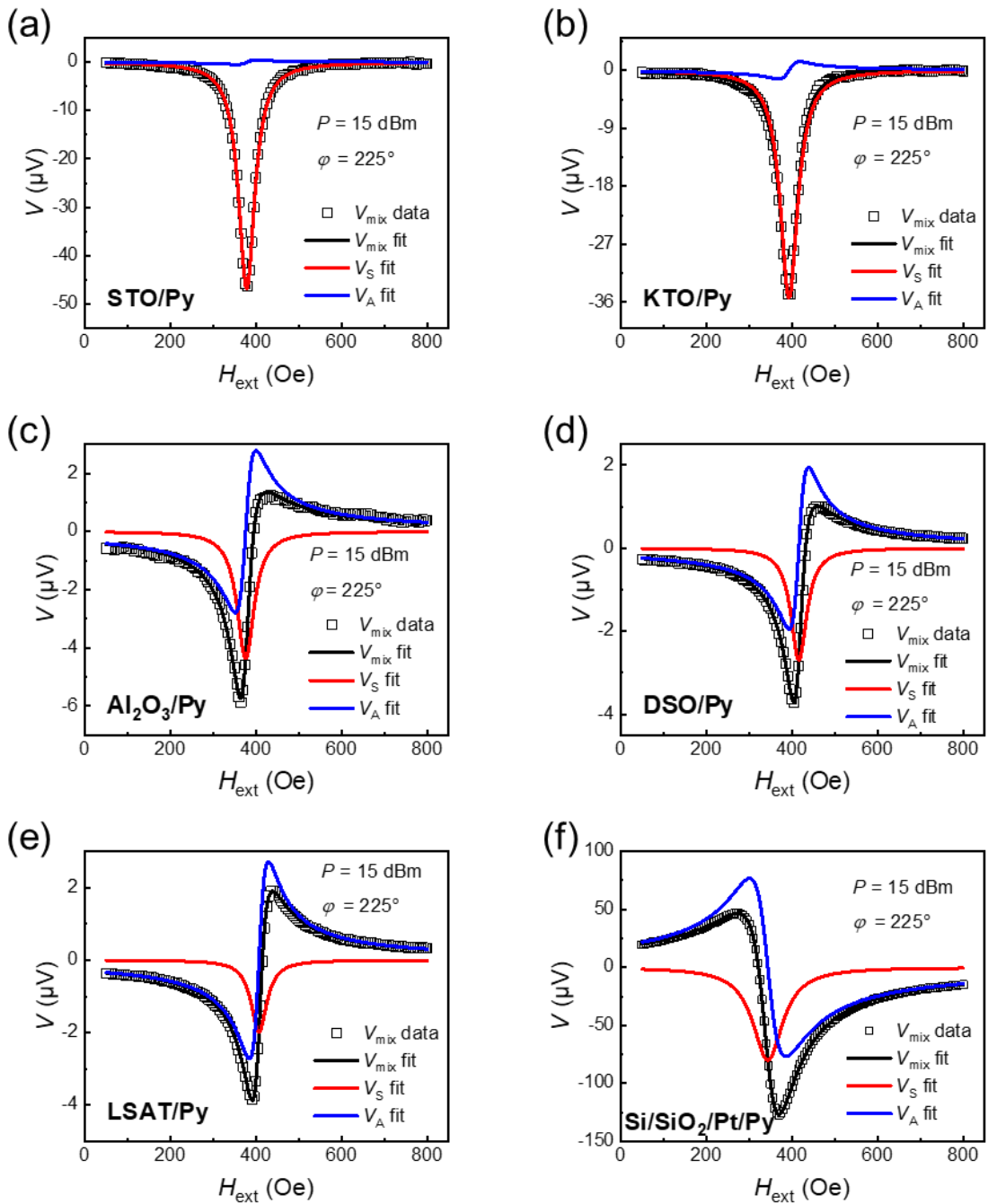

**Fig. 2.** The typical ST-FMR spectra of (a) STO, (b) KTO, (c) Al$_2$O$_3$, (d) DSO, (e) LSAT oxide substrate/Py (5nm)

devices, and (f) Si/SiO$_2$/Pt (5nm)/Py (5nm) device at $f$ = 5 GHz, $\varphi$ = 225°, $P$ = 15 dBm. The black hollow squares represent the experimental data. The black lines, red lines, and blue lines are the sum of symmetric and antisymmetric Lorentz functions fitting $V_{mix}$, symmetric Lorentzian function fitting $V_S$, and antisymmetric Lorentzian function fitting $V_A$, respectively.

To understand the origin of the artifact symmetric voltage signal observed in ST-FMR measurements, we established a revised ST-FMR analysis model by considering the leakage RF current in oxide substrates. As shown in Fig.1, the device can be considered as a resistance $R_{Py}$ of Py and a capacitor $C_{substrate}$ formed between two electrodes that are connected in parallel. At the frequency of ST-FMR measurement (5-10 GHz), the capacitive reactance of $C_{substrate}$ decreases as the frequency increases, and thereby the substrates might become conductive at radio frequencies. As a result, part of the RF current will flow through the oxide substrate due to the current shunting. Therefore, as shown in Fig. 1, the total RF current $I_{RF}$ through the oxide substrate/Py device consists of two components: the RF current $I_{RF-Py}$ through $R_{Py}$, and the RF current $I_{RF-substrate}$ through $C_{substrate}$. Similar to the conventional spin-Hall source material/FM bilayer, $I_{RF-substrate}$ in the oxide substrate will generate an RF Oersted field $H'_{RF}$, as shown in Fig. 3(a). However, the field-like torque of Oersted field would generally give rise to an antisymmetric signal [12], which cannot explain the artifact symmetric signal we observed. It should be noted that there is a 90-degree phase difference between $I_{RF-Py}$ and $I_{RF-substrate}$, as shown in Fig. 3(b). Thereby, due to the 90-degree phase difference between the symmetric and antisymmetric signals, the off-phased $H'_{RF}$ generated from $I_{RF-substrate}$ would cause the appearance of the artifact symmetric signal.

To account for this signal, we revised the ST-FMR analysis model starting with the LLG equation to get the mixing voltage of oxide substrate/Py (see S4 in Supplementary Material [49] for details). The revised mixing voltage is then expressed as Eq. (5):

$$V'_{mix} = -\frac{1}{4}\frac{dR}{d\varphi}\frac{|I_{RF}|\cos\varphi}{\Delta H \, 2\pi df/dH_{ext}|_{H_{ext}=H_0}}\sqrt{1+\left(4\pi M_{eff}/H_0\right)}\frac{\gamma\mu_0|I_{RF-substrate}|}{2w}F_S(H_{ext})$$
$$= V'_s F_S(H_{ext})$$
, (5)

where $w$ is the width of micro stripe, $F_S(H_{ext})$ is the symmetric Lorentzian function, $V'_s$ is the amplitude for that symmetric signal. From Eq. (5), we could conclude that the field-like torque of $H_{RF}^{'} \propto I_{RF,substrate}$ would actually give rise to an artifact symmetric signal $V_{S}^{'}$ in the revised model owing to the 90-degree phase difference between $I_{RF-substrate}$ and $I_{RF-Py}$. As we mentioned earlier, the symmetric signal can only be generated when there is a finite damping-like torque from the spin-Hall source layer in the conventional ST-FMR analysis model (Eq. (1) and Eq. (2)). To link the artifact symmetric signal to the capacitive reactance of substrate, we rewrite $V_S^{'}$ as:

$$V_S^{'} = \left[ -\frac{1}{4}\frac{dR}{d\varphi} \frac{\gamma\mu_0 |P|\cos\varphi}{\Delta H \, 2\pi df/dH_{ext}|_{H_{ext}=H_0}} \left( \sqrt{1+(4\pi M_{eff}/H_0)} \frac{1}{2w} \right) \right] \frac{1}{|\chi_{substrate}|}, \quad (6)$$

where $|P|$ is the modulus of power of the microwave used in ST-FMR measurements which can be expressed as $|P| = |I_{RF}|^2 |Z_{device}|$, and $\chi_{substrate}$ is the reactance of $C_{substrate}$ which can be expressed as $\chi_{substrate} = 1/j2\pi f C_{substrate}$. The part on the left side of $1/|\chi_{substrate}|$ in Eq. (6) can be approximatively regarded as a constant and thereby $V_S^{'}$ would show a linear dependence on $1/|\chi_{substrate}|$.

To experimentally explore the above correlation, we have first measured the impedance of oxide substrate/Py devices and bare oxide substrates by using a network analyzer through the GSG electrodes at frequencies of 5-10 GHz. Due to the parallel connection, the impedance of oxide substrate/Py device $Z_{device}$ can be expressed as Eq. (7):

$$Z_{device} = \frac{R_{Py}}{1+(2\pi f C_{substrate} R_{Py})^2} \left( 1 - j2\pi f C_{substrate} R_{Py} \right). \quad (7)$$

Fig. 3(c) shows the modulus of impedance for oxide substrate/Py devices and bare oxide substrates, along with the phase of impedance of the oxide substrate/Py devices. We found that $|Z_{substrate}|$ ($|\chi_{substrate}|$) decreases as $C_{substrate} \propto \varepsilon_r$ and $|Z_{device}|$ is smaller than $|Z_{substrate}|$ for different oxide substrates, and $\alpha_{device}$ indicated by $\alpha_{device} = \arctan(-2\pi f \cdot R_{Py} C_{substrate})$ being close to -90 degrees as $|Z_{substrate}|$ decreases, which indicates the parrellel connection between $R_{Py}$ and $C_{substrate}$. Especailly, $|Z_{device}|$ becomes close to $|Z_{substrate}|$ for STO and KTO due to their high relative permittivity, suggesting that $I_{RF-substrate}$ would dominant over $I_{RF-Py}$ in STO/Py and KTO/Py devices. Fig. 3(d) summarizes the measured artifact symmetric signals as a function of $1/|\chi_{substrate}|$ for different oxide substrate/Py devices at a frequency of 5 GHz (see Fig. S13 in Supplementary Material [49] for the correlations between $V'_S$ and $1/|\chi_{substrate}|$ at 6-10 GHz). The deviation from the linear fitting of $V'_S$ v.s. $1/|\chi_{substrate}|$ for samples with small permittivity substrates at higher frequencies shown in Fig. S13 may be attributed to some self-induced spin-torque signals from single Py and other spurious effects such as the spin pumping from Py [54,55]. As shown in Fig. 2, given the significant dependence on substrates, other self-induced effects of the single Py are negligible compared with the artifact symmetric signal arising from the leakage RF current in substrate. The observed linear relation between $V'_S$ and $1/|\chi_{substrate}|$ is in good agreement with our model expressed as Eq. (6), providing strong evidence that the artifact symmetric voltage signal in oxide substrate/Py indeed arises from the leakage RF current in the oxide substrate.

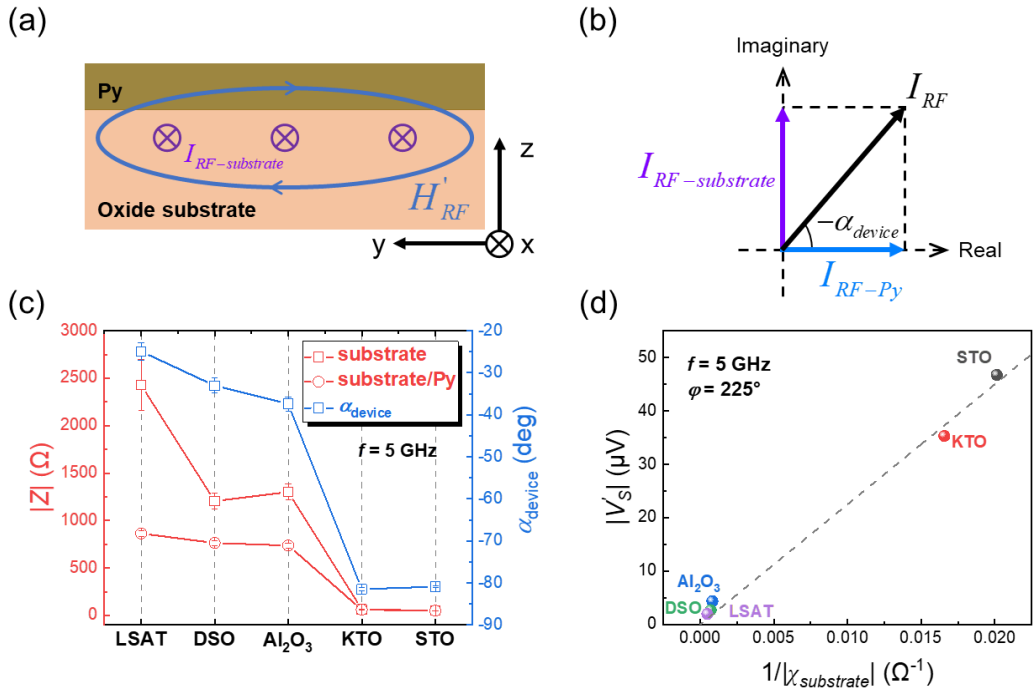

**Fig. 3.** (a) Sectional view of the oxide substrate/Py device. $H'_{RF}$ is the RF Oersted field generated by the leakage RF current $I_{RF-substrate}$ in oxide substrate. (b) The relation between total RF current $I_{RF}$, RF current in oxide substrate $I_{RF-substrate}$, and RF current in Py $I_{RF-Py}$. The length of the arrow represents the modulus of RF current and $\alpha_{device}$ is the phase of impedance of the oxide substrate/Py device. (c) The modulus of impedance $|Z|$ of the different oxide substrates, oxide substrate/Py devices, and the phase of impedance $\alpha_{device}$ of the different oxide substrate/Py devices measured using a vector network analyzer at $f = 5$ GHz, respectively. (d) The correlation between the absolute value of the artifact symmetric voltage signal $V'_S$ and the capacitive reactance of oxide substrate

$\left|\chi_{substrate}\right|$ at $f$ = 5GHz, $\varphi$ = 225°. The dashed line is a linear fit.

To directly evaluated the influence of the artifact symmetric signal from oxide substrates with high relative permittivity on determining the spin-Hall ratio with the ST-FMR measurements, we prepared Pt (5nm)/Py (5nm) and Py (5nm)/Pt (5nm) with a reversed layered sequence on the STO substrate. In these samples, the total RF current $I_{RF}$ consists of $I_{RF-Py}$, $I_{RF-Pt}$ and $I_{RF-substrate}$, in which $I_{RF-Pt}$ gives rise to a symmetric signal $V_S$ from damping-like torque on Py, while $I_{RF-substrate}$ gives rise to the artifact symmetric signal $V_S'$. As shown in Fig. 2, the artifact signal from $I_{RF-substrate}$ has the same symmetry as that from $I_{RF-Pt}$, which means the $V_S'$ and $V_S$ have the same sign when Pt and oxide substrate is on the same side of Py. However, when Pt and oxide substrate are on the opposite side of Py, $V_S'$ and $V_S$ would have an opposite sign and cancel out to each other. Specifically, in the STO/Pt/Py sample, $V_S'$ from STO would add up to $V_S$ from Pt as shown in Fig. 4(a), resulting in an overestimation of $\theta_{SH}$ for Pt; while conversely, in the STO/Py/Pt sample, $V_S'$ would counteract $V_S$, as shown in Fig. 4(b), resulting in an underestimation of $\theta_{SH}$.

Fig. 4(c) shows the spin-Hall ratio $\theta_{SH}$ extracted from the ST-FMR measurements (see S5 in Supplementary Material [49] for details). $\theta_{SH}$ is found to be 0.097 ± 0.001 for STO/Pt/Py, which is about 2.5 times larger than 0.038 ± 0.001 for STO/Py/Pt, consistent with our model. To validate our findings further, we measured the damping-like torque of STO/Pt/Py and STO/Py/Pt by using an independent low-frequency SHHV measurements (see S5 in Supplementary Material [49] for details). In this measurement, the leakage current in oxide substrate should be minimal since the driving frequency for the measurement is low (233 Hz). The damping-like effective fields $H_{DL}$ for these two samples extracted from SHHV were found to be almost the same within experimental uncertainties, as shown in Fig. 4(c). The large discrepancy of the spin-Hall ratios for these two samples using ST-FMR measurements (while almost no difference when using SHHV) demonstrates that substrates with high relative permittivity significantly influence the estimation of the spin-Hall ratio, even in spin-Hall source materials with high conductivity, in ST-FMR measurements.

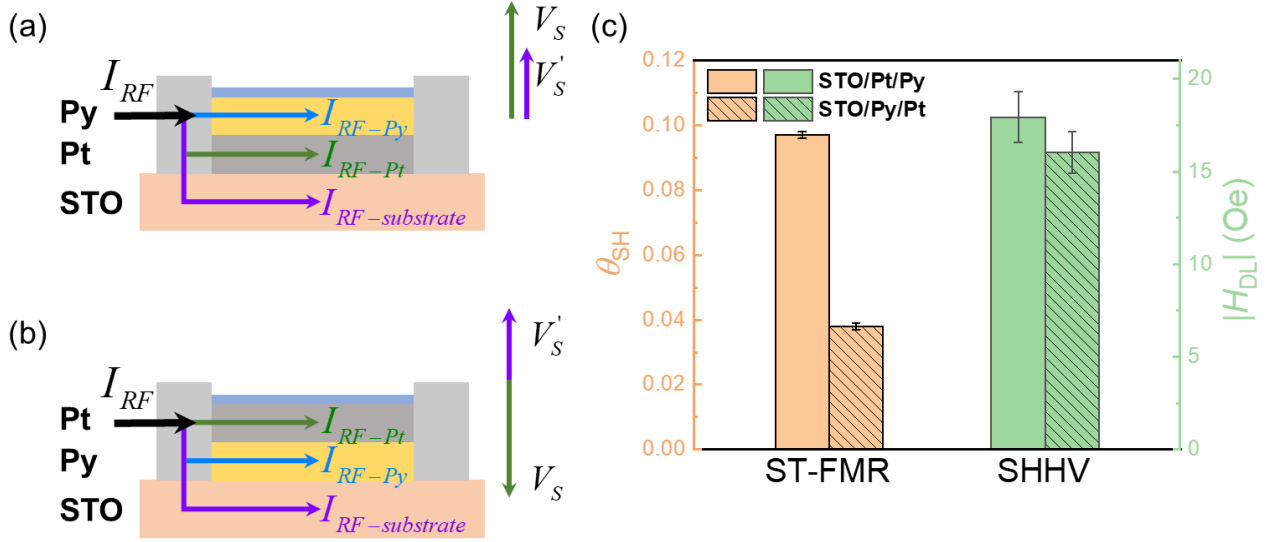

**Fig. 4.** Schematic of the influence of the artifact symmetric voltage on the spin-Hall ratio of Pt in (a) STO/Pt/Py and (b) STO/Py/Pt samples. $I_{RF}$ is the total RF current. $I_{RF-Py}$, $I_{RF-Pt}$ and $I_{RF-substrate}$ are the RF current in Py, Pt and STO substrate, respectively. $V_S$ is the symmetric component related to the damping-like torque from $I_{RF-Pt}$, $V_S^{'}$ is the artifact symmetric signal generated from the leakage RF current $I_{RF-substrate}$ in STO. (c) The spin-Hall ratio $\theta_{SH}$ obtained from ST-FMR measurements, along with the damping-like effective fields $H_{DL}$ obtained from SHHV of STO/Pt/Py and STO/Py/Pt.

## V. Discussion

While the self-calibrated ST-FMR measurement offers unique advantages over other spin torque measurement techniques, it is crucial to recognize that certain substrates with high relative permittivity can introduce significant artifact spin-torque signals in ST-FMR measurement, particularly in spin-Hall source materials with high resistivity. For example, 4d/5d transition metal oxide thin films have recently exhibited exceptionally large spin-Hall ratios [15,28-30,34,37], often possess higher resistivity than heavy metals. When measuring the oxide thin films grown on high-relative-permittivity substrates such as STO and KTO using ST-FMR measurement, their large resistivity can lead to the majority of RF current flowing in substrate, leading to a pronounced artifact symmetric signal.

Previous studies on 4d/5d transition metal oxide thin films and oxide 2DEG systems have frequently used the ST-FMR technique with the analysis of the symmetric signal. Large variations in amplitude and even in sign of spin-Hall ratio have been for the same spin-Hall material with different epitaxial substrates and layered structure [15,28-30,33,37,38,56]. Our results indicate that this varation could mainly be attributed to the artifact symmetric signal caused by using substrates with high relative permittivity (Fig. 3(d)) and the sequence of heterostructures (Fig.4(a) and 4(b)), rather than the other intrinsic mechanisms. For instance, our ST-FMR results for the STO/SIO (17nm)/Py (5nm) sample (see S6 in Supplementary Material [49] for details) show nearly the same magnitude of symmetric signal as compared to the STO/Py (5nm) sample (Fig. 2(a)), indicating that the influence of artifact symmetric signal from STO is overwhelming.

We proposed that employing the dc current-tuned ST-FMR measurements [12,13,15,29] may eliminate the influence of the artifact signal from substrate since there is no obvious correlation between the artifact symmetric signal and the current-induced change of linewidth (see Fig. S7(b) in Supplementary Materials [49]). Furthermore, spin torque measurement techniques at quasi-static frequencies, such as SHHV [29,57,58] or current-induced loop-shift[21-23], should be adopted when characterizing the spin-Hall materials with high resistivity on substrate with high permitivity.

## VI. Conclusions

In conclusion, we have uncovered the presence of a significant artifact symmetric signal in ST-FMR measurements of oxide substrate/Py devices. By revising the conventional ST-FMR analysis model, we have demonstrated that the artifact symmetric signal arises from the leakage RF current in the oxide substrate, caused by the RF current shunting effect. We have further established a correlation between the magnitude of the artifact symmetric signal and the reactance of the capacitor formed in substrate through the GSG electrode. We have evaluated the influence of artifact symmetric signal on the spin-Hall ratio of Pt from oxide substrate with high relative permittivity (i.e., STO) by carrying out the ST-FMR and SHHV measurements, which indicates that the substrates with high relative permittivity significantly affect the evaluation of the spin-Hall ratio in the ST-FMR measurements. Our findings shed light on the critical influence of substrate-induced artifacts in ST-FMR measurements and highlight the importance of usage of suitable measurement techniques when studying spin-Hall source materials.

## Acknowledgments

This work was supported by the National Natural Science Foundation (52161135103, 52250418), and Tsinghua University Initiative Scientific Research Program. W.S. acknowledges grant no. 2021/40/Q/ST5/00209 (Sheng) from the National Science Centre, Poland.

# Supplementary Material

# Artifact from Spin-Torque Ferromagnetic Resonance (ST-FMR) Measurements with Oxide Substrates

Dingsong Jiang[1#], Hetian Chen[2#], Guiping Ji[1], Yahong Chai[1], Chenye Zhang[1], Yuhan Liang[2], Jingchun Liu[2], Witold Skowroński[3], Pu Yu[4], Di Yi[2*], Tianxiang Nan[1*]

1. School of Integrated Circuit and Beijing National Research Center for Information Science and Technology (BNRist), Tsinghua University, Beijing China
2. School of Materials Science and Engineering, Tsinghua University, Beijing China
3. AGH University of Science and Technology, Institute of Electronics, Kraków Poland
4. Department of Physics, Tsinghua University, Beijing, China

\#: The authors contribute equally to this work.

\*: diyi@mail.tsinghua.edu.cn, nantianxiang@mail.tsinghua.edu.cn

# S1. The measurements of the impedance of different oxide substrates and oxide substrate/Py devices.

We used the vector network analyzer (VNA) to measure the $S_{11}$ reflection parameters of bare oxide substrates and oxides substrate/Py devices through the ground-signal-ground (GSG) electrodes at frequencies of 1-15 GHz, as shown in Fig. S1. Subsequently, the impedance of substrate $Z_{substrate}$ and substrate/Py device $Z_{device}$ can be evaluated via $Z = 50\Omega \cdot (1+S_{11})/(1-S_{11})$, where $Z$ and $S_{11}$ are both complex numbers. We calculated the mean value of the 100 data points around each frequency to obtain the $Z_{substrate}$ and $Z_{device}$ at 5, 6, 7, 8, 9, 10 GHz, as shown in Fig. S2 and Fig. S3.

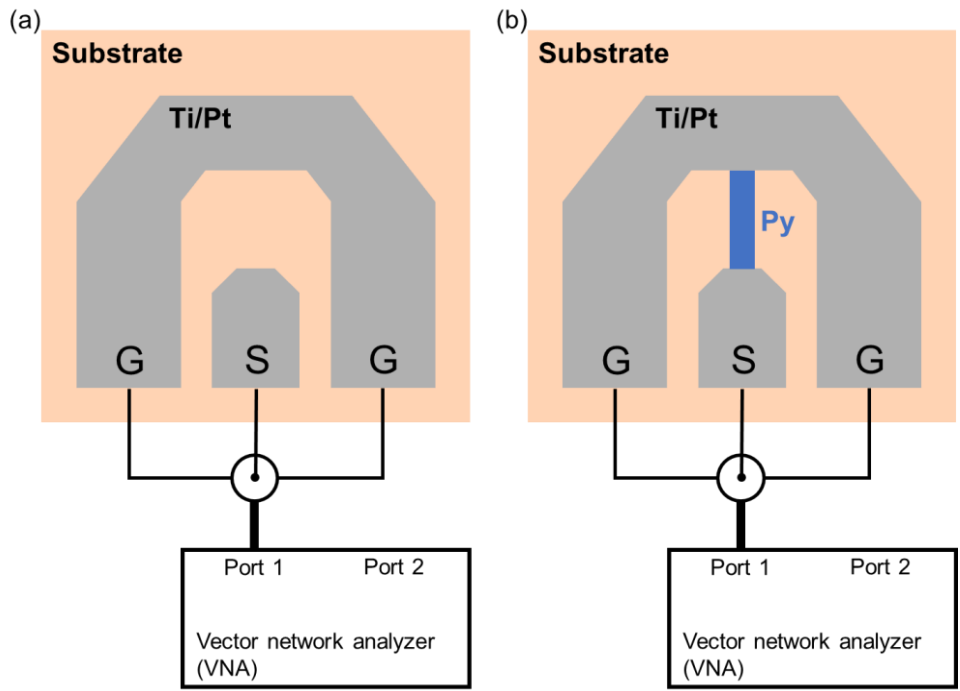

**Fig. S1.** Schematic of using the VNA to measure $S_{11}$ reflection parameter of (a) oxide substrate and (b) oxide substrate/Py device through the ground-signal-ground (GSG) electrodes. When measuring $S_{11}$ reflection parameter of the bare oxide substrate, we only fabricated the GSG electrodes on the substrate.

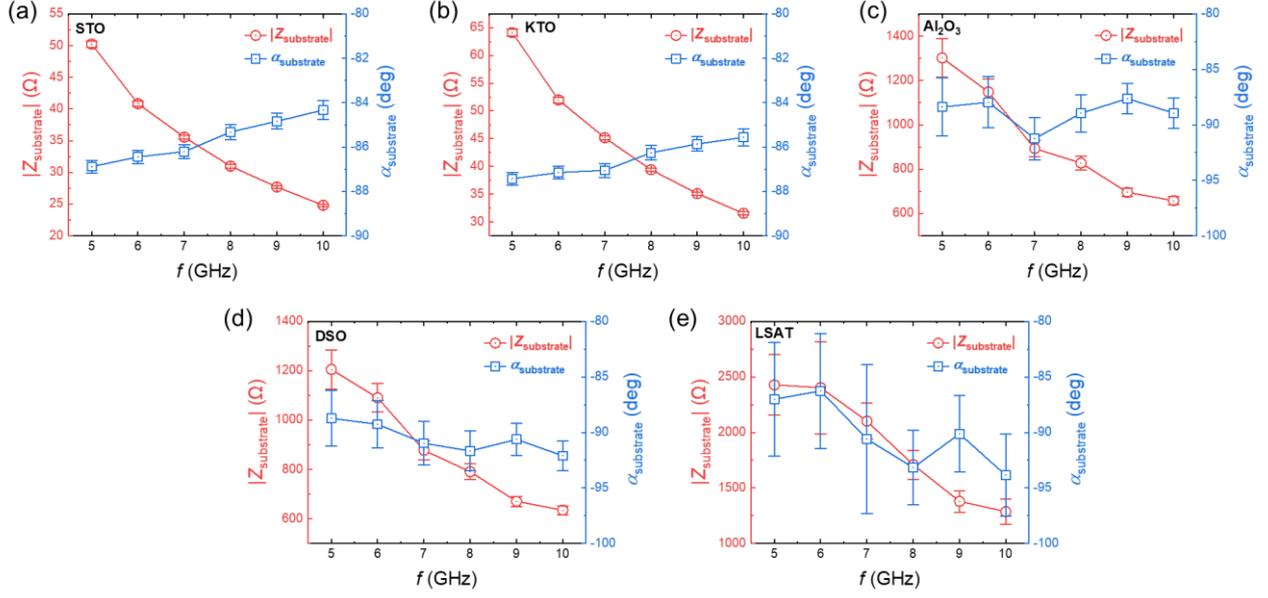

**Fig. S2.** The measured modulus $|Z_{substrate}|$ and phase of impedance $\alpha_{substrate}$ of the different bare substrates at frequencies of 5-10 GHz. $|Z_{substrate}|$ decreases along with the frequency and $\alpha_{substrate}$ is around -90 degrees, which indicates that the capacitor is formed between the GSG electrodes.

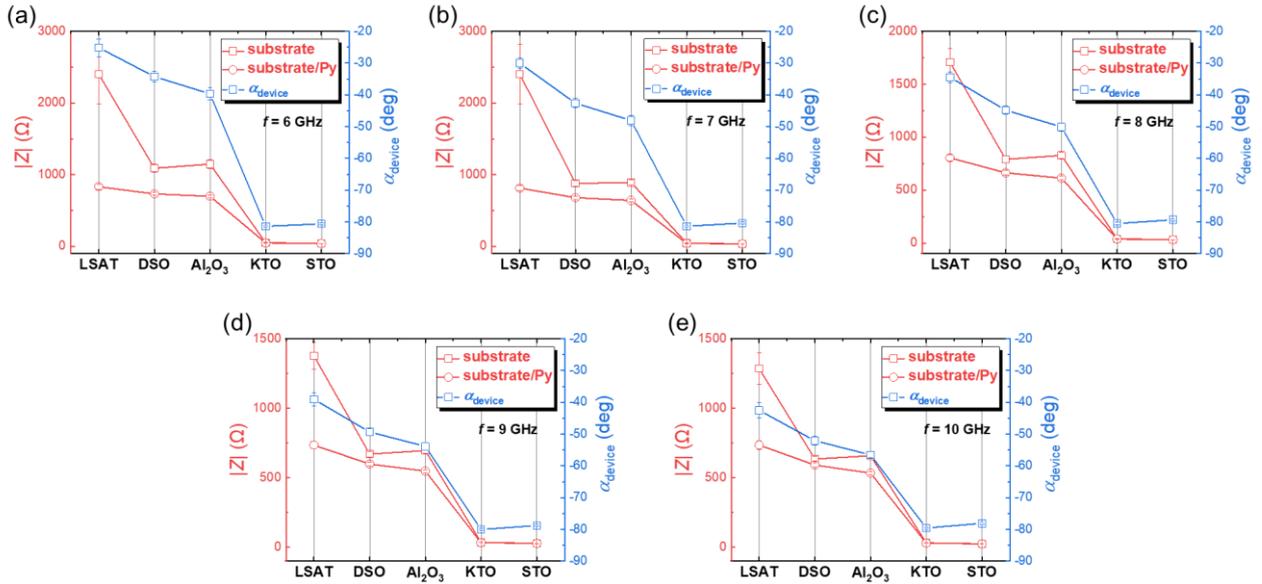

**Fig. S3.** The modulus of impedance of different oxide substrates $|Z_{substrate}|$, oxide substrate/Py devices $|Z_{device}|$ and the phase of impedance of the different oxide substrate/Py devices $\alpha_{device}$ at frequencies of 6-10 GHz.

# S2. Extended results of the ST-FMR measurements of different oxide substrate/Py (5nm) samples and Si/SiO$_2$/Pt (5nm)/Py (5nm) control sample.

We carried out the ST-FMR measurements with varying the frequency to quantify the spin-Hall ratio of Pt in control sample Si/SiO$_2$/Pt (5nm)/Py (5nm). Fig. S4(a) shows the ST-FMR spectra at frequencies of 5-10 GHz, and the symmetric components and antisymmetric components can be extracted via fitting by Eq. (1) in the main text (Fig. S4(b)). We obtained the effective demagnetization of Py by fitting the correlation between resonant frequency $f$ and resonant field $H_0$ via the Kittle formula $f = \gamma/2\pi \sqrt{H_0(H_0 + 4\pi M_{eff})}$, where $\gamma = 0.00294$ GHz/Oe, as shown in Fig. S4(c). The spin-Hall ratio evaluated via $\theta_{SH} = \frac{V_S}{V_A}\sqrt{1+\frac{4\pi M_{eff}}{H_0}}\frac{e\mu_0 M_s t d}{\hbar}$ at frequencies of 5-10 GHz is shown in Fig. S4(d) showing independent on frequency, and the spin-Hall ratio 0.071 ± 0.002 is the mean value.

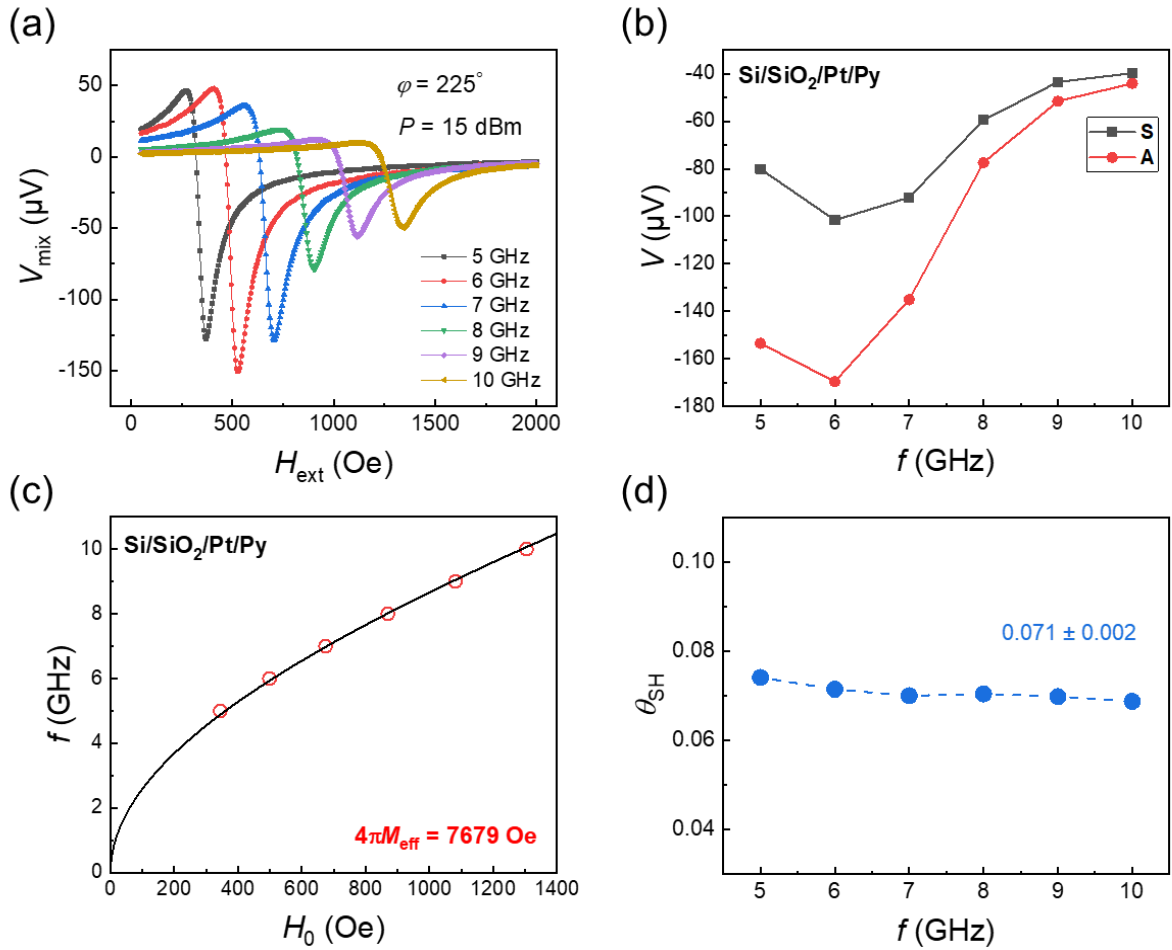

**Fig. S4. ST-FMR measurements of the Si/SiO$_2$/Pt (5nm)/Py (5nm) control sample.** (a) ST-FMR spectra measured at frequencies of 5-10 GHz, $\varphi = 225°$, $P = 15$ dBm. (b) Symmetric and antisymmetric components extracted by the Lorentzian function fitting at frequencies of 5-10 GHz. (c) The resonant frequency $f$ as a function of the resonant

field $H_0$, which be well fit by the Kittle formula. (d) The spin-Hall ratio at frequencies of 5-10 GHz.

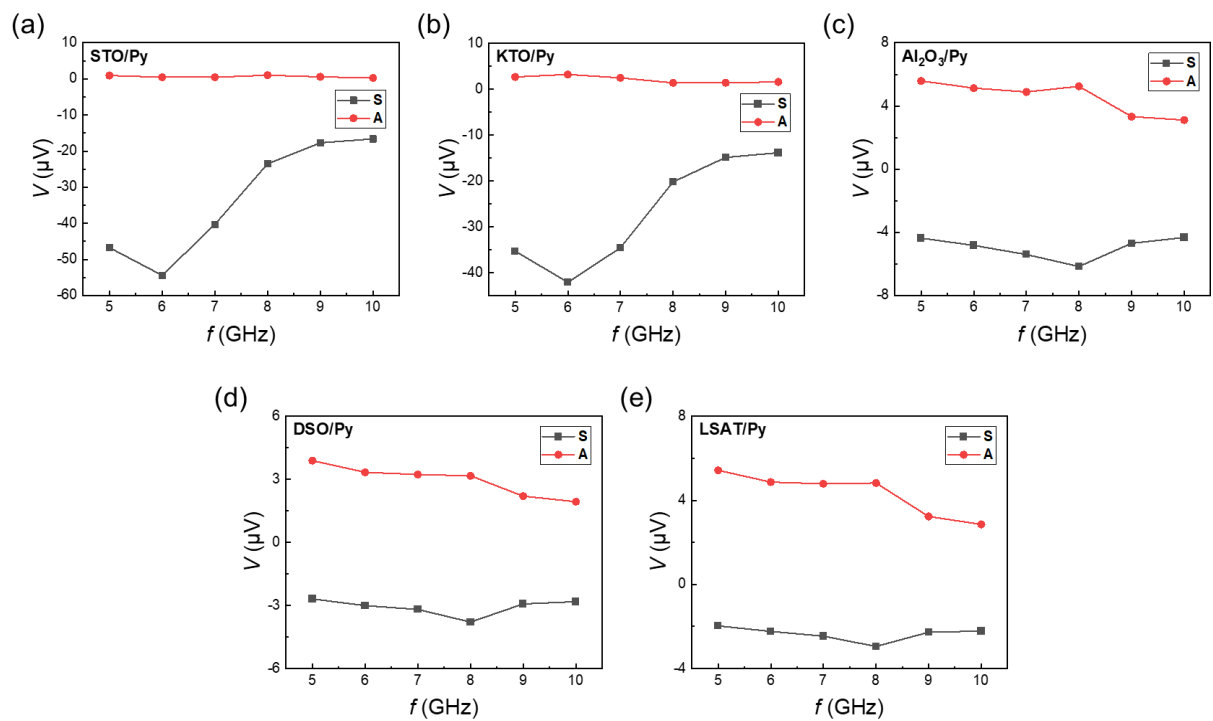

**Fig. S5.** Symmetric and antisymmetric component of (a)-(e) different oxide substrate/Py (5nm) devices extracted by the Lorentzian function fitting at frequencies of 5-10 GHz, $P$ = 15 dBm, and $\varphi$ = 225°.

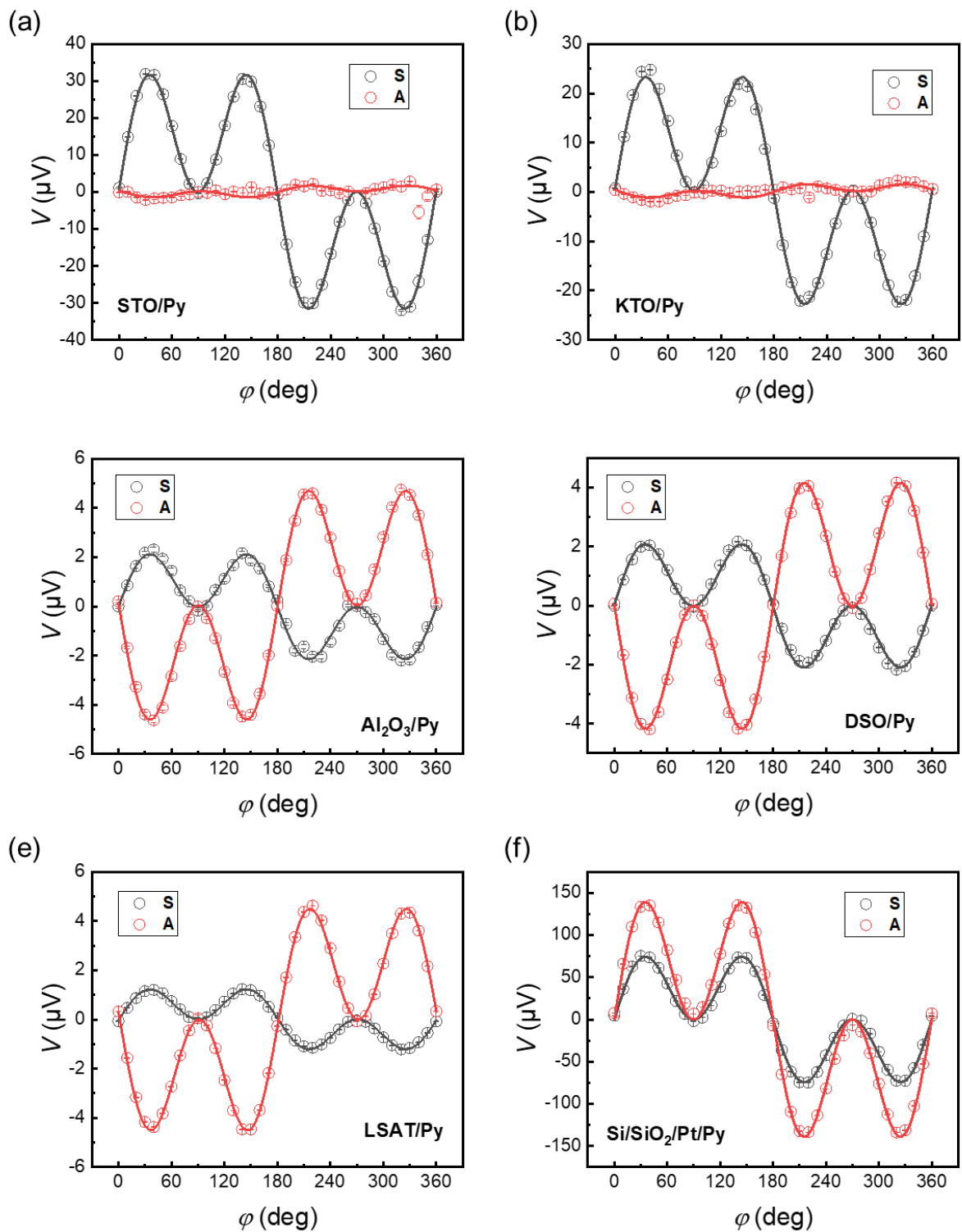

**Fig. S6.** Angular dependence of symmetric components and antisymmetric components of (a)-(e) different oxide substrate/Py devices, and (f) Si/SiO$_2$/Pt (5nm)/Py (5nm) device, which can be well fit by $\sin(2\varphi)\cos(\varphi)$.

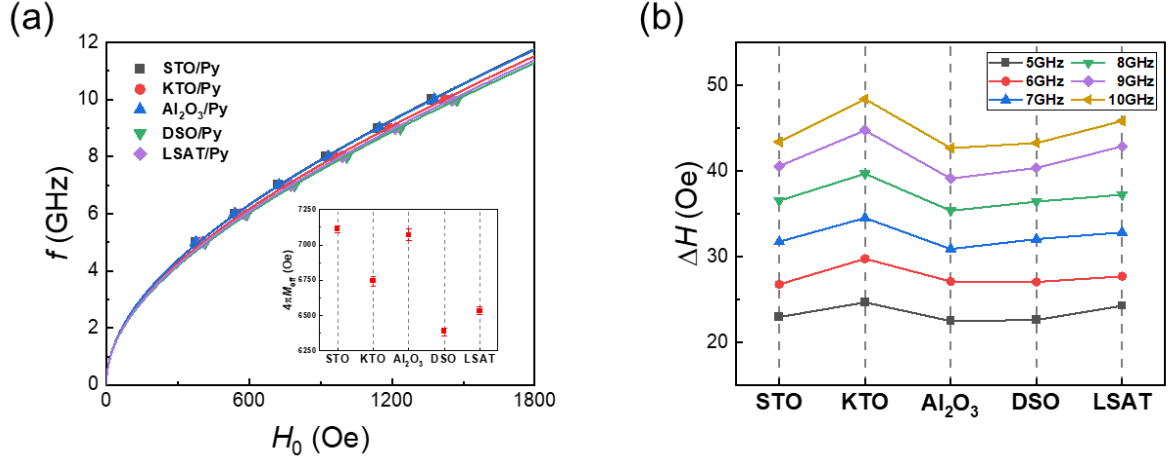

**Fig. S7.** (a) The Kittle formula fitting of resonant frequency $f$ and resonant field $H_0$ of different oxide substrate/Py devices, and inset is the extracted effective demagnetization of Py in different oxide substrate/Py devices. (b) Resonant linewidth of different oxide substrate/Py devices at frequencies of 5-10 GHz, $\varphi = 225°$.

# S3. Absence of damping-like effective field in different oxide-substrate/Py (5nm) samples from in-plane second harmonic Hall voltage measurements.

We used the in-plane second harmonic Hall voltage measurements (SHHV), as shown in Fig. S8(a), to confirm whether there exist damping-like torques in oxide substrate/Py samples. In the SHHV measurements, a 4mA-peak, 233Hz sine current is supplied by a current source, and the first and second harmonic Hall voltage responses are measured by two lock-in amplifiers, respectively. In the macro-spin approximation, the SHHV under an in-plane magnetic field is given by [1]

$$V_{2\omega} = V_a \cos\phi + V_p \cos 2\phi \cos\phi + V_T \sin 2\phi, \quad (S1)$$

where $\phi$ is the angle of $H_{in}$ with respect to the current direction, $V_a = V_{AHE} H_{DL}/2(H_{in} + H_k) + V_{ANE}$ is the contribution from damping-like field $H_{DL}$, and anomalous Nernst effect $V_{ANE}$ due to the vertical thermal gradient, $V_{AHE} = IR_{AHE}$ is the anomalous Hall effect and $R_{AHE}$ is the anomalous Hall effect resistance, $V_p = V_{PHE}(H_{FL} + H_{Oe})/H_{in}$ is the contribution from field-like field $H_{FL}$ and Oersted field $H_{Oe}$, $V_{PHE} = IR_{PHE}$ is the planar Hall effect and $R_{PHE}$ is the planar Hall effect resistance, $V_T$ is the contribution from a thermal gradient along the direction of current [1]. Fig. S8(b)-(e) show the typical SHHV results of STO/Py, KTO/Py, Al$_2$O$_3$/Py, DSO/Py, LSAT/Py under a 1000 Oe magnetic field, which can be well fit by Eq. (S1). To extract $H_{DL}$, we have determined $R_{AHE}$ of different oxide substrate/Py devices under the swept out-of-plane field $H_z$, as shown in Fig. S9. Here, $4\pi M_{eff}$ from the Kittle fitting in Fig. S7(a) is employed to replace $H_k$. The linear correlation between $V_a$ and $-V_{AHE}/2(H_{in}+4\pi M_{eff})$ is shown in Fig. S10(a), and the main component $V_a$ is mostly contributed by anomalous Nernst effect since $V_a$ is hardly dependent of exerted magnetic field. The extracted damping-like fields $H_{DL}$ which are in the range of -0.18 ± 0.21 Oe to 0.12 ± 0.23 Oe for these five types of oxide-substrate/Py devices, as shown in Fig. 10(b), can be ignored within the error range.

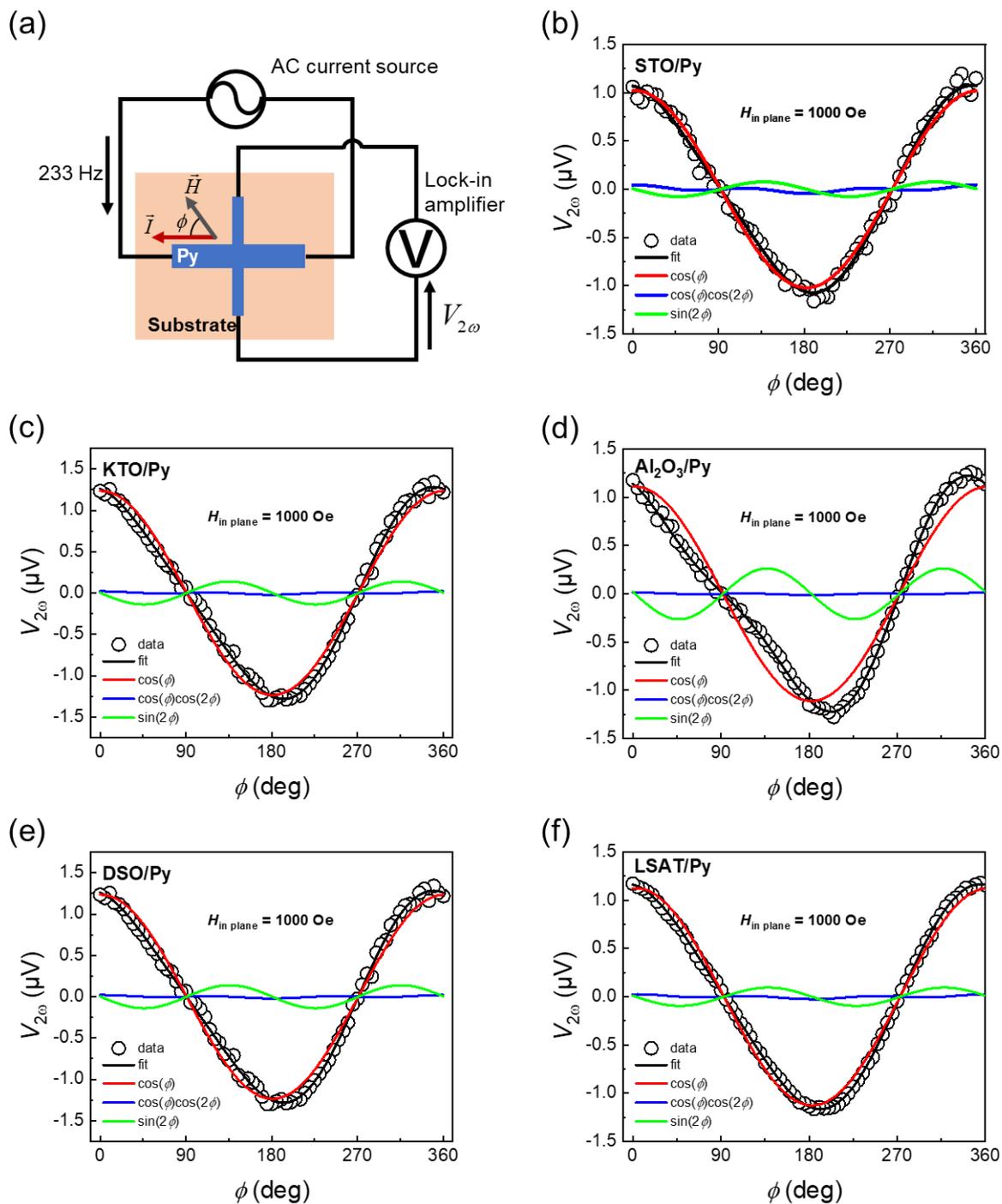

**Fig. S8. SHHV measurements of different oxide substrate/Py devices.** (a) Schematic of in-plane second harmonic Hall voltage measurement set-up. (b)-(e) The typical in-plane second harmonic Hall voltage responses under a 1000 Oe magnetic field of different oxide substrate/Py devices, which can be well fit by Eq. (S1).

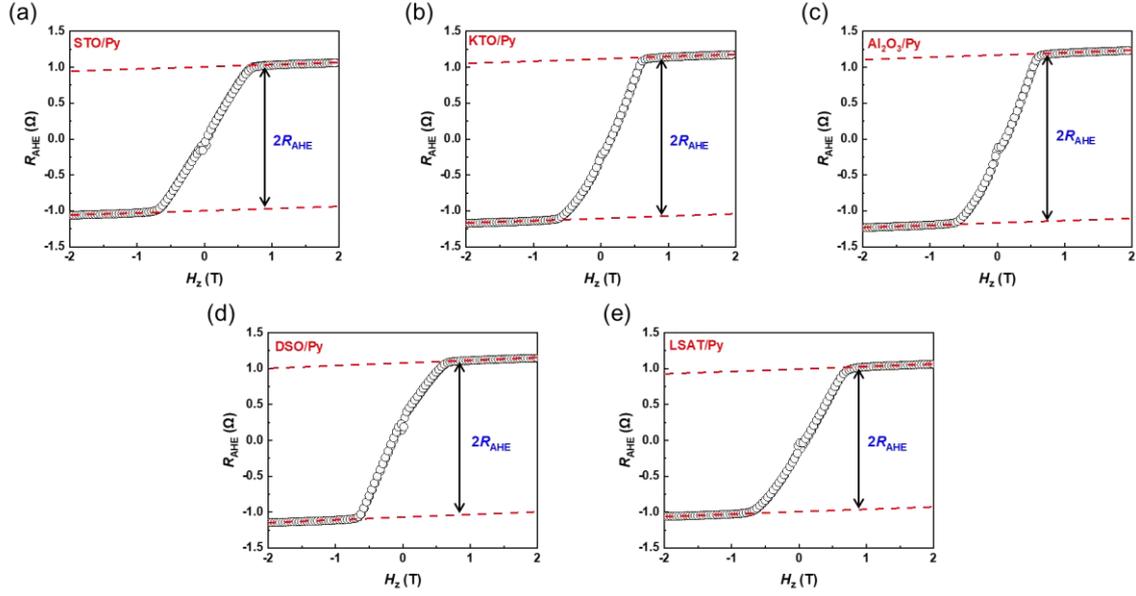

**Fig. S9.** The resistance of anomalous Hall effect $R_{AHE}$ under a swept of out-of-plane field $H_z$ at range of -2 T to 2 T of different oxide substrate/Py Hall bars.

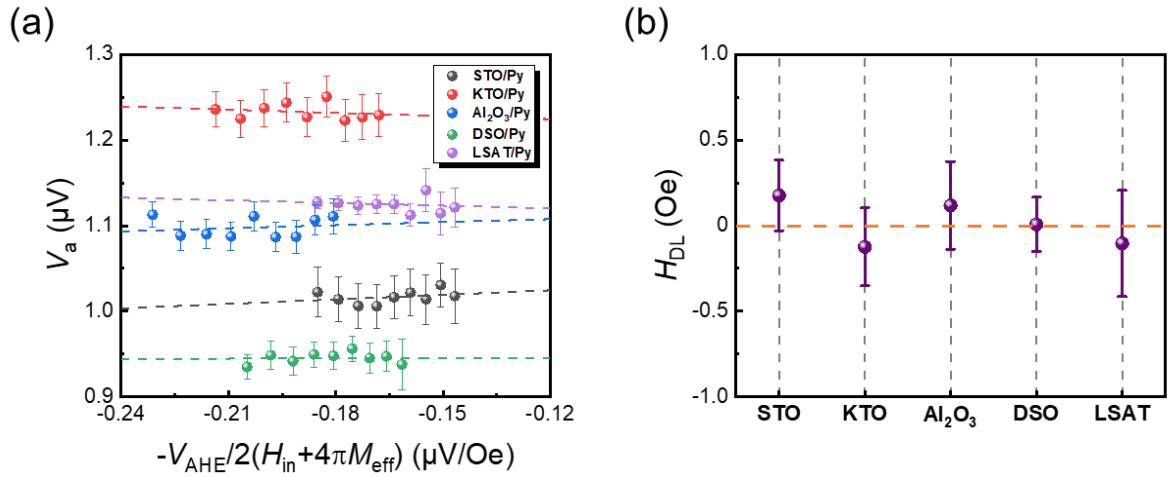

**Fig. S10.** (a) $V_a$ as a function of $-V_{AHE}/2(H_{in}+4\pi M_{eff})$, with the slopes being damping-like effective field $H_{DL}$. (b) $H_{DL}$ extracted from SHHV measurements of different oxide substrate/Py devices. Within the margin of error, $H_{DL}$ of different oxide substrate/Py devices can be ignored.

# S4. The derivation of mixing voltage of the oxide substrate/Py samples from ST-FMR.

By solving the Landau-Lifshitz-Gilbert (LLG) equation

$$\frac{d\vec{m}}{dt} = -\gamma\mu_0 \vec{m} \times \vec{H}_{eff} + \alpha \vec{m} \times \frac{d\vec{m}}{dt} + \tau_{DL}\vec{m} \times (\vec{\sigma} \times \vec{m}) + (\tau_{FL} + \gamma\mu_0 H_{RF})\vec{m} \times \vec{\sigma}, \quad (S2)$$

where $\gamma$ is the gyromagnetic ratio, $\alpha$ is the Gilbert damping coefficient, $\mu_0$ is the permeability in vacuum, $\tau_{DL}$ is the damping-like torque, $\tau_{FL}$ is the field-like torque, $H_{eff}$ is the sum of external field $H_{ext}$ and the demagnetization field $4\pi M_{eff}$, $H_{RF}$ is the RF Oersted field, $\vec{m}$ is the magnetization of FM layer, $\vec{\sigma}$ is the spin polarization, we can get the magnetization component of FM layer [2]:

$$m_y = \frac{1}{2} \frac{\cos\varphi}{2\pi(df/dH_{ext})|_{H_{ext}=H_0}} \frac{(\tau_{FL} + \gamma\mu_0 H_{RF})\sqrt{1+(4\pi M_{eff}/H_0)} + i\tau_{DL}}{H_{ext} - H_0 + i\Delta H}, \quad (S3)$$

where $\varphi$ is the angle between RF current and external field, $f$ is the frequency of ST-FMR measurement, $H_0$ is the resonant field, $M_S$ is the saturation magnetization, $H_k$ is the magnetic anisotropy field, $\Delta H$ is the linewidth.

The mixing voltage signal is

$$V_{mix} = \frac{1}{2}\frac{dR}{d\varphi}I_{RF}\text{Re}(m_y), \quad (S4)$$

where $I_{RF}$ is the RF current through micro strip, $R$ is the resistance of microstrip.

For the oxide substrate/Py device, the leakage RF current $I_{RF-substrate}$ in the oxide substrate will generate an RF Oersted field $H'_{RF}$, as shown in Fig. 3(a) in the main text. To evaluate the distribution of RF current in substrate, we did finite element simulations to investigate the distribution of RF current in the substrate. We focused solely on the RF current along the x-axis which generates an Oersted field along the y-axis, resulting in the production of an artifact symmetric signal. In our simulations, we introduced an RF power of 32 mW (~15 dBm) at 5 GHz. The material parameters utilized in the simulations are detailed in Table S1, and the simulation model is depicted in Fig. S11(a). Table. S1. The material parameters used in simulations

| Parameter | Value |
| --- | --- |
| The resistivity of Pt electrode | $2.0 \times 10^{-7}$ Ω·m |
| The resistivity of Py | $4.5 \times 10^{-7}$ Ω·m |
| The relative permittivity of substrate | 10, 20, 100, 200, 300 |

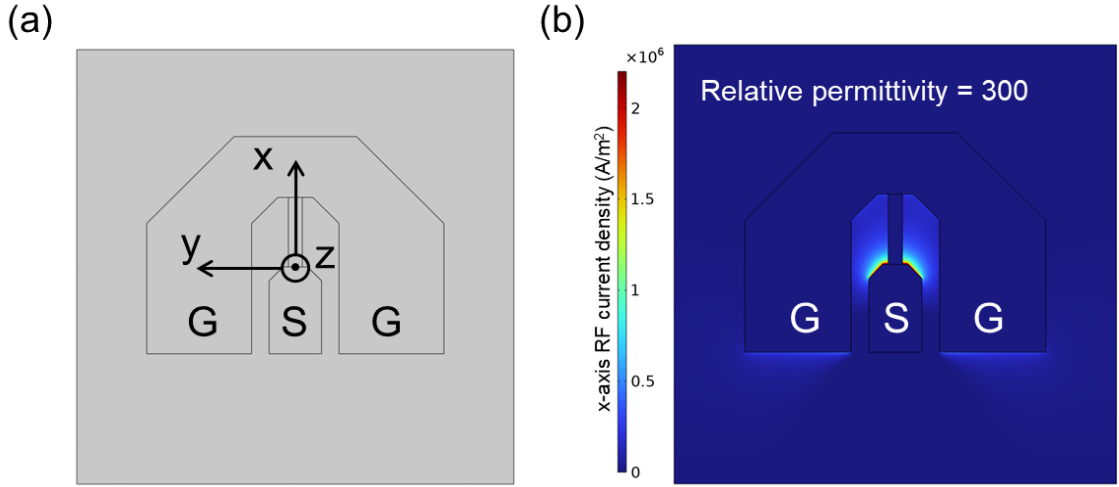

**Fig. S11.** (a) The simulation model of device used in our simulations. (b) The typical simulation results of the RF current distribution.

As shown in Fig. S11(b), due to the spreading effect, the distribution of the x-axis RF current in substrate is wider than the microstrip width, and the RF current is mostly distributed in the vicinity of the interface between the substrate and Py. According to the skin effect, employing 37% of the maximum RF current density as a threshold allows us to evaluate the spreading width of the x-axis RF current distribution, as depicted in Fig. S12(a). When using substrates with relative permittivity ranging from 10 to 300, the width of x-axis RF current distribution is found to be 28.8μm to 29.4μm. With an order of magnitude increase in the relative permittivity of the substrate, there is a concomitant order of magnitude increase in the artifact signal, as shown in Fig.2. Consequently, we think that the primary factor affecting the magnitude of the artifact signal is the disparity in relative permittivity among different substrates.

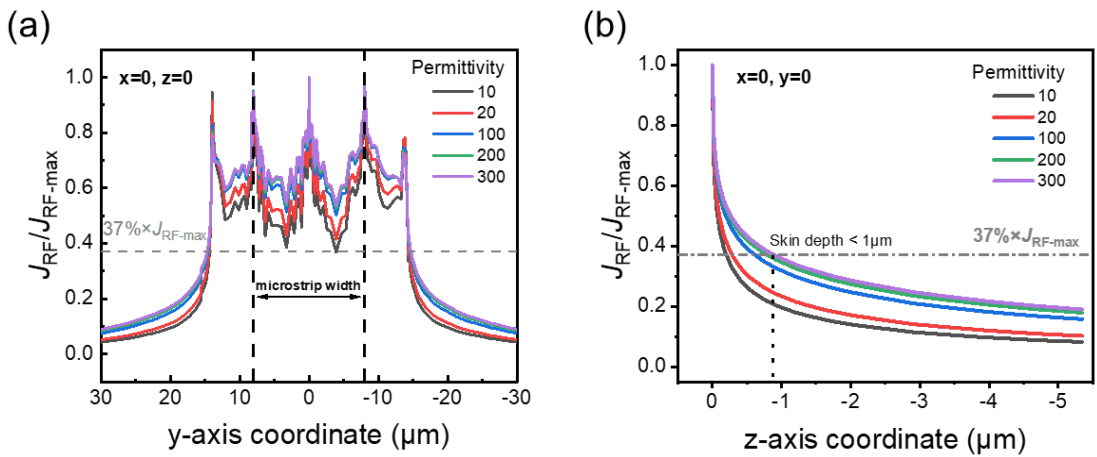

**Fig. S12.** The simulation results of (a) the y-axis distribution of x-axis RF current in the interface between substrate and Py at x=0, z=0, and (b) the z-axis distribution of x-axis RF current in substrate at x=0, y=0.

Moreover, the impact of RF current flowing in the substrate outside the width of the microstrip on the Py layer should be disregarded, as the region where the y-axis component of the Oersted field is present extends beyond the width of the microstrip, and the z-axis component of the Oersted field generated by RF current flowing in substrate on either side of the microstrip cancels out each other. Therefore, the spreading effect of RF current in substrate could be ignored, and only the RF current flowing beneath the microstrip should be considered. As shown in Fig. S12(b), the width of microstrip (~18 μm) is one order larger than the skin depth (<1 μm) of RF current in substrate, so the Oersted field can be determined by the Ampere's law approximately [3]. Considering the nonuniform distribution of leakage RF current in the skin depth, the Oersted field generated from RF current could be expressed in an integral form:

$$H'_{RF} = \int_{-d}^{0} \frac{i|J_{RF-substrate}(z)|dz}{2} = \frac{1}{2}i\int_{-d}^{0} \frac{\frac{|I_{RF-substrate}(z)|}{w}dz}{2} = \frac{i}{2w}\sum |I_{RF-substrate}(z)|, \quad (S5)$$

where $d$ is the skin depth of RF current in substrate, $w$ is the width of microstrip, $\sum I_{RF-substrate}(z)$ is the sum of RF current in the skin depth. In Eq. (S5), $\sum I_{RF-substrate}(z)$ can be replaced with $|I_{RF-substrate}|$ approximately, since the main distribution of RF current in substrate is in skin depth. Thus, $H'_{RF}$ could be expressed approximately as Eq. (S6):

$$H'_{RF} = \frac{1}{2w}i\sum |I_{RF-substrate}(z)| \approx i\frac{|I_{RF-substrate}|}{2w}. \quad (S6)$$

Since there is no spin source material in the oxide substrate/Py, the part associated with $\tau_{DL}$ in Eq. (S3) could be deleted. Meanwhile, the field-like torque could also be ignored in oxide substrate/Py devices, which can be determined from the SHHV measurements as shown in Fig. S8. Substituting Eq. (S6) into Eq. (S3), we can get:

$$m_y = \frac{1}{2}\frac{\cos\varphi}{2\pi(df/dH_{ext})|_{H_{ext}=H_0}} \frac{i\sqrt{1+(4\pi M_{eff}/H_0)}\cdot\gamma\mu_0 \frac{|I_{RF-substrate}|}{2w}}{H_{ext}-H_0+i\Delta H}, \quad (S7)$$

Furthermore, we can get the mixing voltage in oxide substrate/Py from Eq. (S4) and Eq. (S7):

$$V'_{mix} = -\frac{1}{4}\frac{dR}{d\varphi}\frac{|I_{RF}|\cos\varphi}{\Delta H 2\pi(df/dH_{ext})|_{H_{ext}=H_0}}\sqrt{1+(4\pi M_{eff}/H_0)}\frac{\gamma\mu_0|I_{RF-substrate}|}{2w}F_S(H_{ext})$$
$$= V'_S F_S(H_{ext}) \quad (S8)$$

where $F_S(H_{ext}) = \Delta H^2/[\Delta H^2 + (H_{ext}-H_0)^2]$ is the symmetric Lorentzian function.

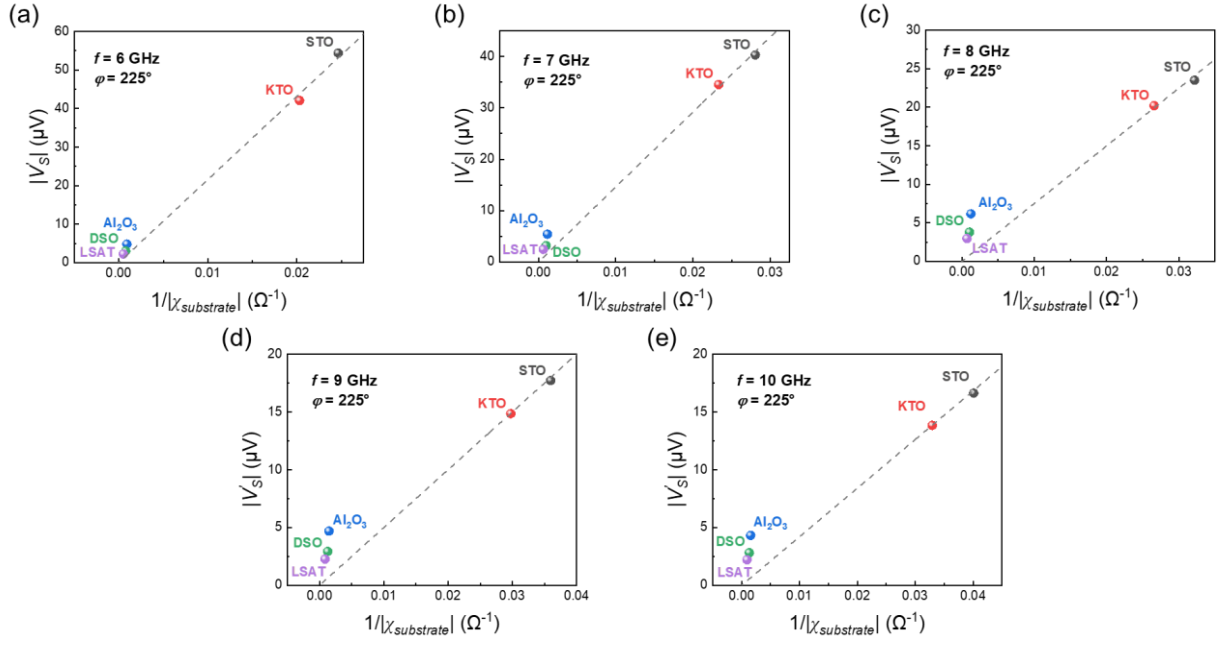

**Fig. S13.** The linear correlation between the absolute value of artifact symmetric signal $|V'_S|$ and $1/|\chi_{substrate}|$ at frequencies of 6-10 GHz, $\varphi = 225°$. The dashed line is the linear fit.

# S5. ST-FMR and SHHV measurements of STO/Pt (5nm)/Py (5nm) and STO/Py (5nm)/Pt (5nm).

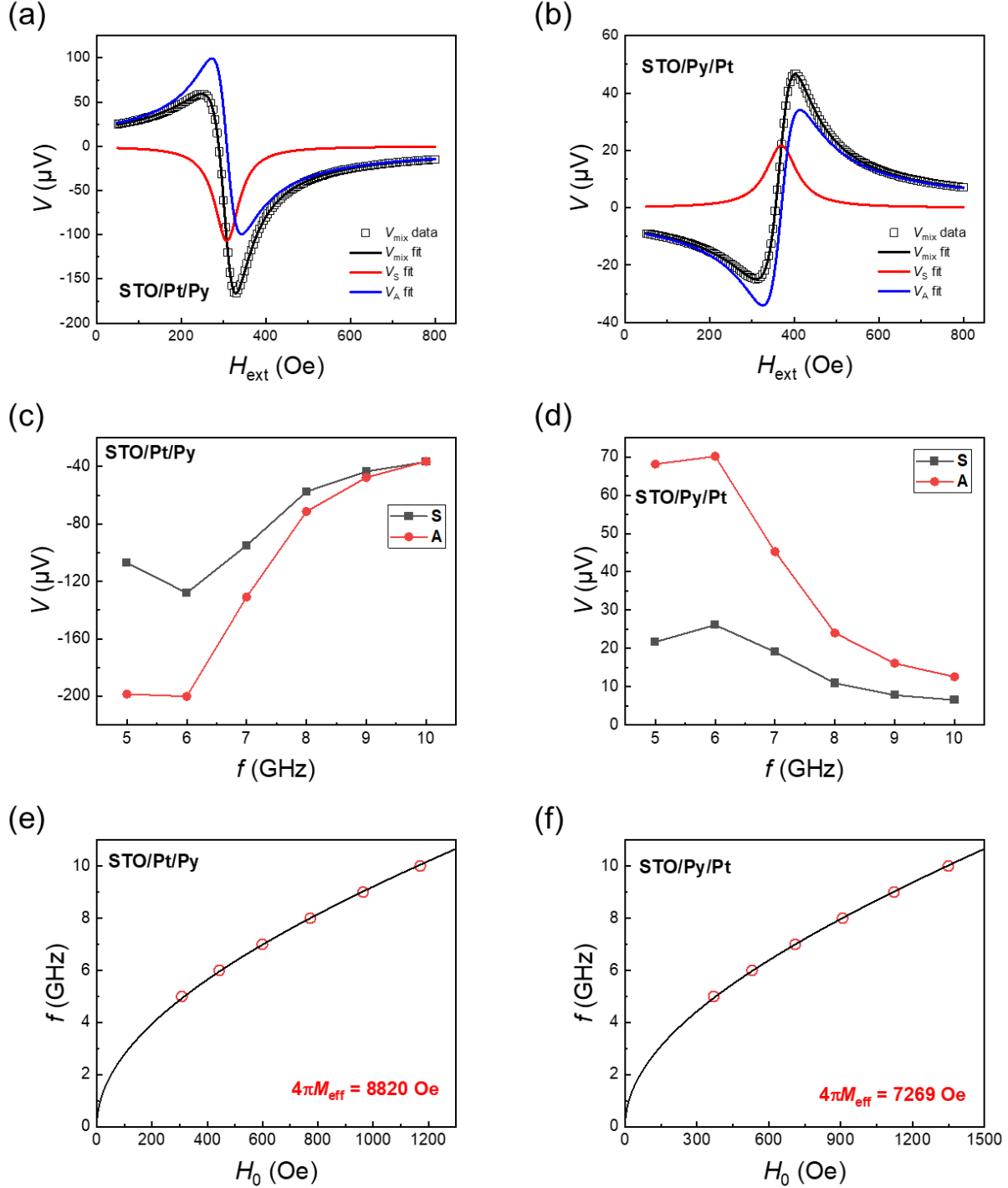

**Fig. S14. ST-FMR measurements of STO/Pt (5nm)/Py (5nm) and STO/Py (5nm)/Pt (5nm).** Typical ST-FMR spectra measured at a frequency of 5 GHz, $\varphi$ = 225°, $P$ = 15 dBm of (a) STO/Pt/Py and (b) STO/Py/Pt, which can be well fit via Eq. (1) in the main text. Symmetric and antisymmetric components of (c) STO/Pt/Py and (d) STO/Py/Pt extracted by the Lorentzian function fitting at frequencies of 5-10 GHz. The Kittle formula fitting of resonant

frequency *f* and the resonant field $H_0$ of (e) STO/Pt/Py and (f) STO/Py/Pt.

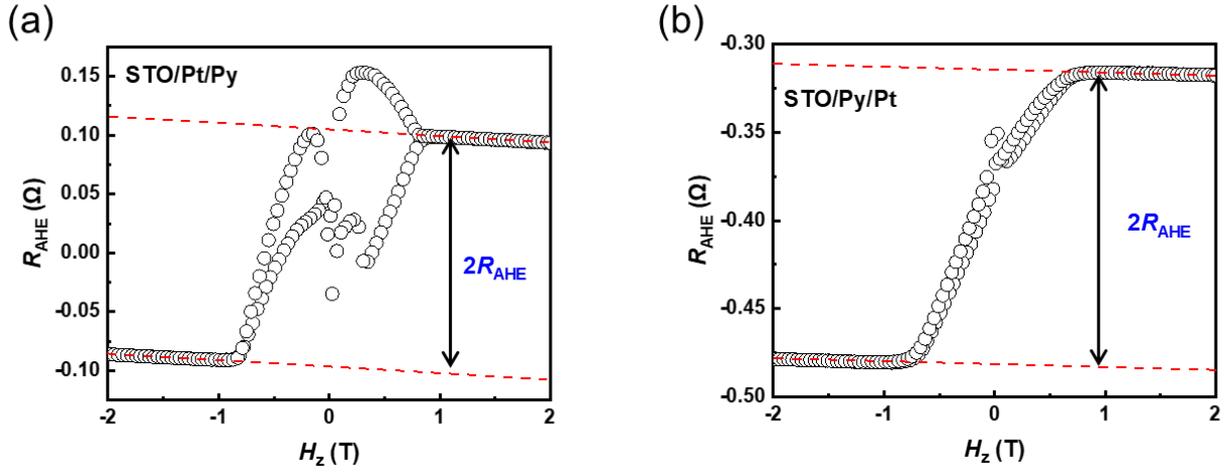

**Fig. S15.** The resistance of anomalous Hall effect $R_{AHE}$ under a swept of out-of-plane field $H_z$ at range of -2 T to 2 T of (a) STO/Pt/Py and (b) STO/Py/Pt Hall bars.

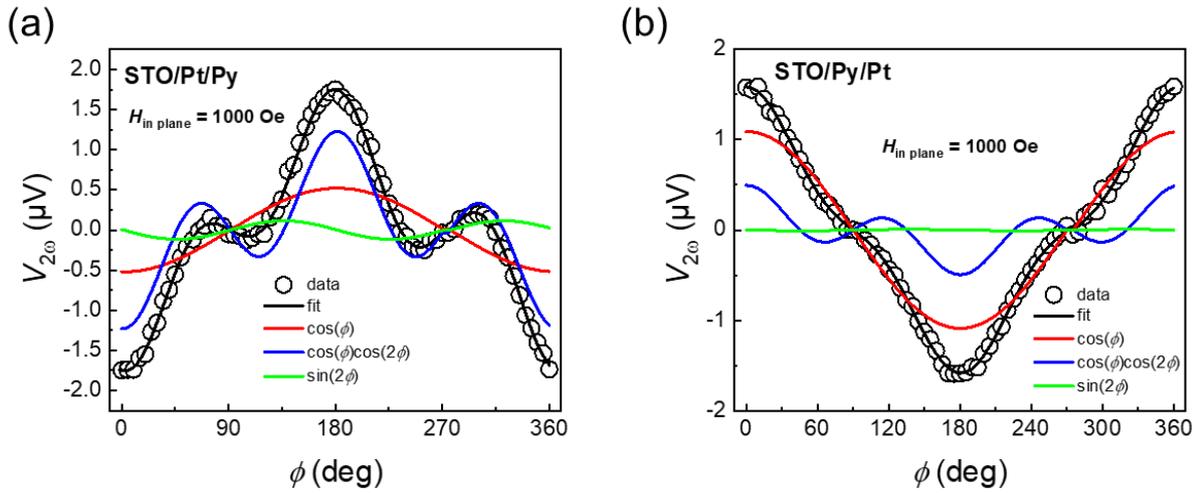

**Fig. S16.** The typical in-plane second harmonic Hall voltage responses under a magnetic field of 1000 Oe of (a) STO/Pt/Py and (b) STO/Py/Pt, which are well fit by Eq. (1). During the SHHV measurements of STO/Pt/Py and STO/Py/Pt samples, an 8mA-peak, 233Hz sine current is supplied by a current source.

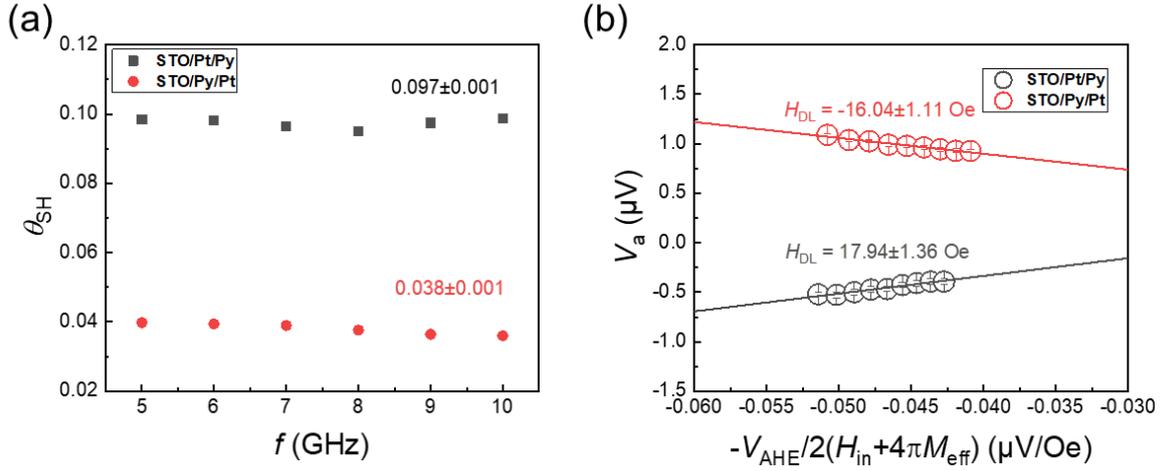

**Fig. S17.** (a) The spin-Hall ratio $\theta_{SH}$ of STO/Pt/Py and STO/Py/Pt evaluated $\theta_{SH} = \dfrac{V_S}{V_A}\sqrt{1+\dfrac{4\pi M_{eff}}{H_0}}\dfrac{e\mu_0 M_s t d}{\hbar}$ from ST-FMR measurements at frequencies of 5-10 GHz. (b) The linear dependence of $V_a$ on $-V_{AHE}/2(H_{in}+4\pi M_{eff})$ of STO/Pt/Py and STO/Py/Pt, and the slopes being damping-like effective field $H_{DL}$

## S6. ST-FMR results of the STO/SIO (17nm)/Py (5nm) sample.

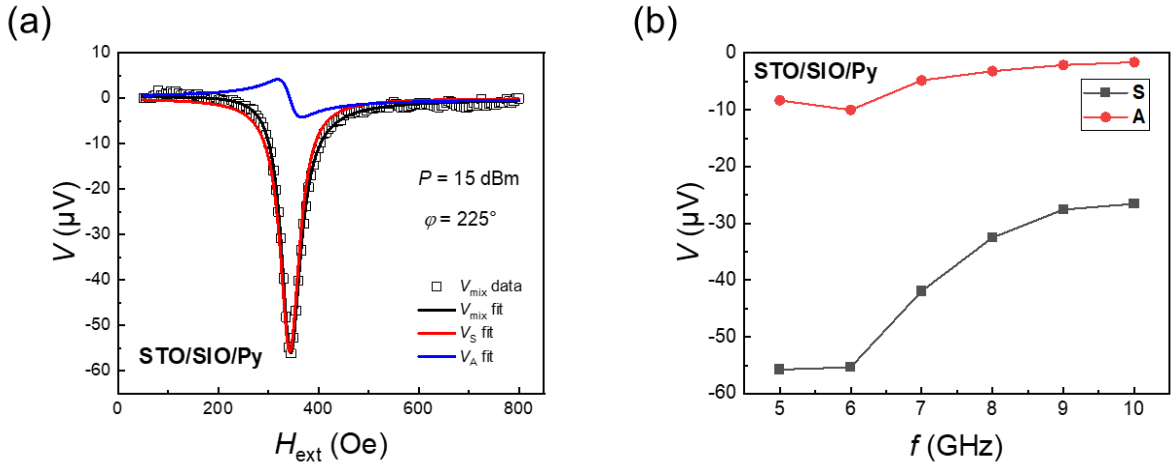

**Fig. S18. ST-FMR measurements of STO/SIO (17nm)/Py (5nm) sample.** (a) Typical ST-FMR spectra of STO/SIO/Py at $f$ = 5 GHz, $\varphi$ = 225°, and $P$ = 15 dBm. (b) Symmetric and antisymmetric components extracted by the Lorentzian functions fitting at frequencies of 5-10 GHz.

## References.